\documentclass[%
 reprint,
 amsmath,amssymb,
 aps,
]{revtex4-1}

\usepackage{graphicx}
\usepackage{dcolumn}
\usepackage{bm}
\usepackage{amsmath}
\usepackage{multirow}
\usepackage[version=3]{mhchem}
\usepackage[dvipsnames]{xcolor}
\usepackage{pifont}
\usepackage[linktocpage=true,
  colorlinks=true, 
  pdfborder={0 0 0},
  linkcolor=blue,
  citecolor=red,
  filecolor=yellow,
  urlcolor=blue,
  bookmarks,
  pdfauthor={},
]{hyperref}

\begin{document}

\preprint{APS/123-QED}

\title{Comment on the stability of decorated C$_{48}$B$_{12}$ heterofullerene}

\newcommand{\Basel}{Department of Physics, Universit\"at Basel,
Klingelbergstr. 82, 4056 Basel, Switzerland}
\newcommand{\Grenoble}{Univ. Grenoble Alpes, INAC-MEM, L\_Sim, F-38000 Grenoble, France}

\author{Deb Sankar De} \affiliation{\Basel}
\author{Santanu Saha} \affiliation{\Basel}
\author{Luigi Genovese} \affiliation{\Grenoble}
\author{Stefan Goedecker} \affiliation{\Basel}





\date{\today}

\newcommand{\saha}[1]{\textcolor{red}{#1}}
\newcommand{\raj}[2]{\textcolor{blue}{#2}}

\begin{abstract}
A good hydrogen storage material should adsorb hydrogen in high concentrations and with optimal binding energies. Numerous mixed carbon boron fullerenes which are decorated with metal atoms were previously constructed by hand and proposed as a promising material in this context. We present a fully ab-initio, unbiased structure search in the configurational space of decorated C$_{48}$B$_{12}$  and find that most of the hitherto postulated ground state structures are not ground states. We determine the energetically lowest configurations for  Be, Ca, Li and Sc decorated C$_{48}$B$_{12}$ clusters. 
\end{abstract}

\pacs{Valid PACS appear here}
\maketitle


\section{\label{sec:level1}Introduction}


The emergence of green energies and new technologies in the past decades has led to a strong growth in research activities in material discovery. Amidst the theoretically and experimentally studied materials, fullerene and fullerene based structures ~\cite{yildirim2005molecular,chandrakumar2008alkali,Kawazoe,Yoon_Ca-metalfunctional-PRL2008}  play a prominent role due to their applicability in various fields. Recent experimental observation of borospherene ~\cite{zhai2014observation}  has generated renewed interest in application of fullerene based structures in the field of green energy.
 

Among the fullerene based structures, the boron substituted fullerene C$_{48}$B$_{12}$  was predicted theoretically by Manna et. al ~\cite{riad2003predicted}. This structure was constructed by substituting 12 C atoms by B atoms homogeneously in the fullerene such that there are no B-B bonds. This structure will be referred to as "diluted" C$_{48}$B$_{12}$ in the following. However, in an unbiased minima hopping (MH) structure prediction  Mohr. et. al. found a patched configuration, 
where all the B atoms are close together in a two-dimensional patch, that is considerably lower in energy than the homogeneously distributed arrangement   \cite{mohr2014boron}. This structure will be referred to as a "patched" configuration. Similar studies have been conducted for decorated Si$_{20}$ ~\cite{Willand2010}  and  exohedrally decorated borospherene (B$_{40}$) ~\cite{saha2017}.

Based on DFT calculations, metal decorated C$_{48}$B$_{12}$ 
has recently been predicted to have superior H$_{2}$
adsorption capabilities compared to pure C$_{48}$B$_{12}$. 
Through a systematic search, Sun et. al. found that that 12 Li atoms that are 
homogeneously distributed on the surface of the fullerene form the most stable configuration~\cite{sun2009functionalized}. Gao et al. constructed two microporous frameworks consisting of
organic linkers and exohedral metallofullerene nodes. They showed through the calculation that the both frameworks can store H$_{2}$ with
a gravimetric density up to 8---9.2 wt \% \cite{gao2014designs}. The patched configurations was not considered as a basic unit in this study. Zhao et. al. reported that a structure, that is homogeneously coated with 12 Sc atoms, is stable and also suitable for H$_{2}$ adsorption but
no structural stability search  was performed~\cite{zhao2005hydrogen}. Er et. al. found that homogeneously 
distributing 6 Ca atoms on the fullerene is stable with respect to decomposition into the fullerene molecule and Ca bulk metal~\cite{er2015improved}. Most recently a similar study was conducted by Qi et al. on the same 6 Ca atom decorated diluted configuration to study the hydrogen uptake mechanism ~\cite{qi2015theoretical}.  Lee et. al constructed a few hand made structures by adding 6 diluted Be atoms on the surface and they concluded that homogeneously distributed Be atoms on C$_{48}$B$_{12}$ give rise to the most stable configuration ~\cite{lee2010ab}.

In these studies of metal decorated C$_{48}$B$_{12}$, two major factors were overlooked: (a)  only diluted structures were decorated with different elements whereas the 
patched structure, that is lower in energy by 1.8 eV, was not considered and (b) the investigated structures were obtained through chemical intuition and local  relaxation. Without knowing the true ground state of the these systems, it has been conjectured that they are good for hydrogen adsorption. A thorough unbiased search of the PES is necessary to obtain reliable conclusions on their stability and their resulting capacity for hydrogen storage.

Using the Minima Hopping Method (MHM) ~\cite{goedecker2004minima}, 
we performed a comprehensive structural search, decorating both diluted and patched C$_{48}$B$_{12}$ fullerenes. Almost in every case (except for Li), the ground state  structure was massively distorted by the decoration and the metal decorated patched configurations are much lower in energy than their diluted counter part.  This is in contrast to C$_{60}$ whose structure remains stable under decoration as was previously shown~\cite{De2017}.  

\section{Computational Methodology:}
All the geometry optimizations 
and the energy calculations have been 
performed on the density functional
theory (DFT) level as implemented in BIGDFT ~\cite{genovese2008daubechies}. This code uses Daubechies wavelets
as a basis set and the
exchange correlation  was described by the Perdew-
Burke-Ernzerhof (PBE) functional ~\cite{perdew1996generalized,Libxc}. A grid spacing of 0.4 Bohr was used. Convergence parameters of BIGDFT
were set such that total energy differences were converged
up to 10$^{-4}$ eV and all configurations were relaxed until
the maximal force component of any atom reached the
noise level of the calculation, which was of the order of
1 meV/~\AA. Soft Goedecker-type pseudopotentials with non-linear core corrections  were used ~\cite{hartwigsen1998relativistic}.
The minima hopping method (MHM) was used for an
unbiased search of new configurations~\cite{goedecker2004minima,schaefer2015stabilized}. It efficiently finds low-energy structures by exploiting the Bell-Evans-Polanyi
principle for molecular dynamics ~\cite{roy2008bell}. The potential energy surface (PES) is explored by performing
consecutive short molecular dynamics escape steps followed by local geometry relaxations.

\section{Results:}


We first adsorbed single atoms on both diluted and patched configurations. There are 8 possible sites to adsorb a single atom on the diluted configuration and 10 on  the patched configuration. We have considered all of of them and report the largest adsorption energy in Table~\ref{tab:bondlength}. 
The adsorption energy of n decorating atoms on C$_{48}$B$_{12}$ is calculated by the following equation:
\begin{equation}
E_{ads}=E_{C_{48}B_{12}}+n\times E_{atoms}-E_{n_{atom}C_{48}B_{12}}
\end{equation} 
For the diluted configuration, single Be, Ca and Sc atoms prefer the hexagonal site which has two neighboring boron atoms. A single  Li atom prefers the hexagonal site where the hexagon has a single boron atom. For the patched configuration, Be atoms prefer the pentagon site with 3 neighboring borons, Li prefers the hexagon site with one boron, Ca prefers the hexagon site with two borons and Sc prefers to be on top of boron when boron is in the middle of a hexagonal site with four neighboring boron atoms. 



\begin{table}[t]
\centering
\caption{~ Largest adsorption energies (E$_{ads}$) for different types of decorating atoms}
\begin{tabular}{|l|l|l|l|l|}
\hline
\multirow{2}{*}{\begin{tabular}[c]{@{}l@{}}Type of\\ Structure\end{tabular}} & \multicolumn{4}{l|}{E$_{ads}$ in eV} \\ \cline{2-5} 
                                                                             & Li     & Be     & Ca    & Sc    \\ \hline
Patch                                                                      &  3.20      &  3.41      &   3.62    &    5.28   \\ \hline
Dilute                                                                       &   2.88     &    2.66    &   2.89    &   4.22    \\ \hline
\end{tabular} \label{tab:bondlength}
\end{table}

\begin{figure}[b!]
\includegraphics[angle=0,width=90mm,scale=0.2]{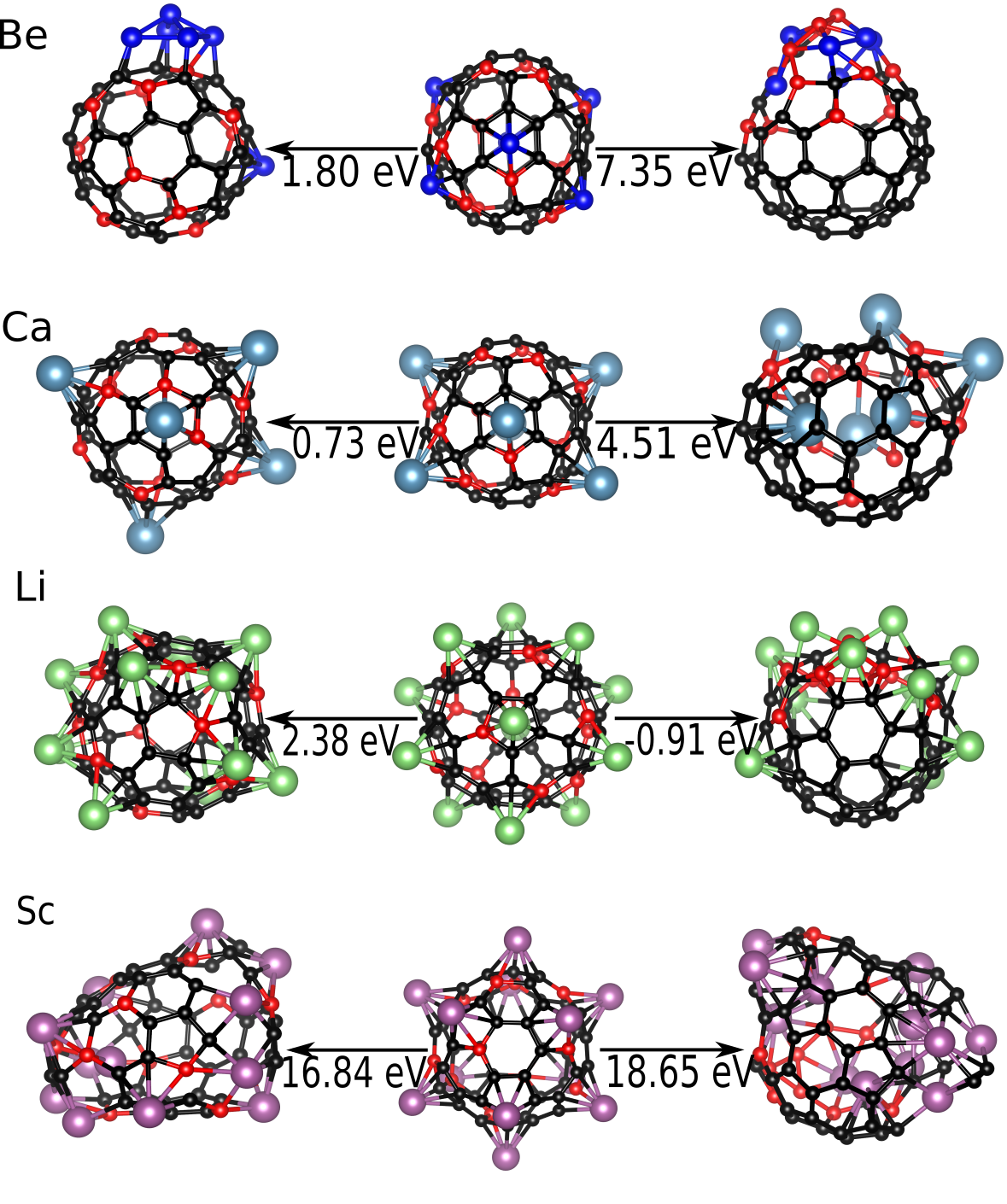}
\caption{The Lowest energy structures for diluted and patched configurations for Be, Ca, Li and Sc decorated C$_{48}$B$_{12}$. The column in the center shows the previously predicted structures. The column on the right  
the lowest energy structures for decorations of the  diluted fullerene and the 
left column the lowest energy structures for the patched fullerene.
The energy difference is given with respect to the middle structure for each type of decoration. It shows that for all types of atoms except Li the  patched configurations are the most stable ones.  }
\label{fig:All_atom}
\end{figure}

We next present our results for the cases where the  C$_{48}$B$_{12}$ fullerene 
is decorated with several metal atoms, namely 12 for Li and Sc and 6 for Ca and Be. Fig.~\ref{fig:All_atom} summarizes the results.


\subsection{Be}

A structure consisting of six Be atoms which are homogeneously distributed on C$_{48}$B$_{12}$ was proposed as a good H absorber ~\cite{lee2010ab}. MH runs revealed that even for the diluted fulerene cage there are many different structures which are lower in energy than this previously proposed one. We observed that the most of the Be atoms form  lusters and that a single Be atom prefers to stay on an hexagonal-site far from the Be-cluster. The underlying diluted fullerne structure is however not destroyed by the decoration.    

Even lower energy configurations can be obtained if one decorates the patched fullerene structure. The patched configuration is broken during the MH runs and a few of the Be atoms go inside the C$_{48}$B$_{12}$. This suggests that though the patched configuration is more stable than the diluted configuration, it is more reactive to Be atoms.



\subsection{Ca}
The hitherto known lowest energy structure consists of 6 homogeneously distributed Ca atoms on the H-sites of the diluted configuration~\cite{er2015improved} ~\cite{qi2015theoretical}.  We found a 
new structure based on a diluted  fullerene that is 0.73 eV lower than the previously predicted  configuration. In contrast to  C$_{60}$, where the 
Ca atoms form a patch~\cite{De2017} they remain homogeneously distributed on the dilute C$_{48}$B$_{12}$. 
The patched configuration with clustered Ca atoms is however also in this case 
by far the lowest in energy. In these configuration some Ca atoms again enter into the cage of the patched fullerene.

\subsection{Sc}
Zao et al proposed a configuration where 12 Sc atoms are homogeneously distributed over surface of the dilute C$_{48}$B$_{12}$ as the most stable configuration~\cite{zhao2005hydrogen}. 
For the Ca decorated  dilute  C$_{48}$B$_{12}$ configuration, we have again observed several lower energy configurations than the previously predicted configuration. 
In the lowest energy configuration the diluted C$_{48}$B$_{12}$ the cage structure is broken and a few Sc atom went inside the cage, resulting in a drastic energy lowering.  

Similarly, in the patched C$_{48}$B$_{12}$ configuration decorated with Sc, the cage gets partially destroyed. This configuration is again the lowest one.

  


\subsection{Li}

12 Li atoms regularly distributed on the P sites of the diluted C$_{48}$B$_{12}$ structure were proposed to be most stable configuration ~\cite{sun2009functionalized}.  
The energy can however be lowered by distributing the Li atoms in a more random way, resulting in some Li-Li bonds. This was the lowest energy configuration found in our study. Decorations of the patched C$_{48}$B$_{12}$ fullerene 
resulted in this case in higher energy configurations.
So decoration with Li atoms stabilizes the diluted configuration.  



\section{\label{sec:level4}Conclusion:}

In conclusion, we have shown that the  ground state of different metal decorated
boron-carbon heterofullerenes is fundamentally different from
previously published structures ~\cite{er2015improved,sun2009functionalized,zhao2005hydrogen,er2015improved,lee2010ab}. 
The metals considered in our study 
are not able to homogeneously coat the dilute C$_{48}$B$_{12}$ cage. 
With the exception of Li, the decoration actually destroys the cage. 
The bonding character in all these configurations is quite complicated and there is no simple rule
available to predict the type of distribution of different elements on the C$_{48}$B$_{12}$ surface.
Our results give also valuable information for experimental synthesis efforts of such heterofullerenes.
Our investigation reveals that most previous computational studies on metal atom decorated fullerenes for hydrogen storage are based on unrealistic structures. 

\bibliographystyle{apsrev4-1}
\bibliography{C48B12}

\end{document}